\documentclass[useAMS,usenatbib,usegraphicx]{mn2e}

\title[White Dwarfs with $K$-Band Excess]{A Trio of Metal-Rich Dust and Gas Disks Found Orbiting
Candidate White Dwarfs with $K$-Band Excess}

\author[J. Farihi et al.]{J. Farihi$^1$\thanks{E-mail: jf123@star.le.ac.uk}, 
B. T. G\"ansicke$^2$, P. R. Steele$^3$, J. Girven$^2$, M. R. Burleigh$^1$,
\newauthor  E. Breedt$^2$ and D. Koester$^4$\\
$^1$Department of Physics  Astronomy, University of Leicester, Leicester LE1 7RH, UK\\
$^2$Department of Physics, University of Warwick, Coventry CV4 7AL, UK\\
$^3$Max Planck Institut f\"ur Astrophysik, D-85741 Garching, Germany\\
$^4$Institut f\"ur Theoretische Physik und Astrophysik, University of Kiel, 24098 Kiel, Germany}

\begin{document}

\date{}

\maketitle

\label{firstpage}

\begin{abstract}

This paper reports follow-up photometric and spectroscopic observations, including warm {\em Spitzer} 
IRAC photometry of seven white dwarfs from the SDSS with apparent excess flux in UKIDSS $K$-band 
observations.  Six of the science targets were selected from 16\,785 DA star candidates identified either 
spectroscopically or photometrically within SDSS DR7, spatially cross-correlated with $HK$ detections in 
UKIDSS DR8.  Thus the selection criteria are completely independent of stellar mass, effective temperature 
above 8000\,K, and the presence (or absence) of atmospheric metals.  The infrared fluxes of one target are
compatible with a spatially-unresolved late M or early L-type companion, while three stars exhibit excess
emissions consistent with warm circumstellar dust.  These latter targets have spectral energy distributions
similar to known dusty white dwarfs with high fractional infrared luminosities (thus the $K$-band excesses).
Optical spectroscopy reveals the stars with disk-like excesses are polluted with heavy elements, denoting
the ongoing accretion of circumstellar material.  One of the disks exhibits a gaseous component -- the fourth
reported to date -- and orbits a relatively cool star, indicating the gas is produced via collisions as opposed 
to sublimation, supporting the picture of a recent event.  The resulting statistics yield a lower limit of 0.8\% 
for the fraction dust disks at DA-type white dwarfs with cooling ages less than 1\,Gyr.  Two overall results 
are noteworthy: all stars whose excess infrared emission is consistent with dust are metal-rich; and no 
stars warmer than 25\,000\,K are found to have this type of excess, despite sufficient sensitivity.
\end{abstract}


\section{INTRODUCTION}

Near-infrared photometry of white dwarfs examines a scientific phase space that is difficult or impossible 
to probe for main-sequence stars.  The first to recognize this potential was \citet{pro81}, who realized the 
compact nature of white dwarfs allowed a straightforward photometric detection of very cool stellar and 
substellar companions.  Brown dwarfs, which often require sophisticated detection techniques when 
searched for at main-sequence stars, are up to 10 times larger than a typical white dwarf and can be 
readily detected in {\em spatially-unresolved} observations as excess near-infrared flux \citep{pro83}.  
Thus, white dwarfs are an excellent tool for studies of the low-mass stellar and substellar mass function 
via companions \citep{far05,zuc92,pro82}.  Photometry with {\em Spitzer} has extended this potential to 
include closely orbiting brown dwarfs and massive planets although no candidates are yet apparent 
\citep{far08b,mul07}.

\begin{table*}
\begin{center}
\caption{Multi-Wavelength Photometry and Derived Parameters for SDSS White Dwarfs\label{tbl1}} 
\begin{tabular}{@{}rccccccc@{}}
\hline\hline

SDSS Prefix					&0959			&1159$^a$		&1221			&1247			&1320			&1506			&1557\\

\hline

Spectral Type					&DAZ			&DQp			&DAZ			&DA				&DA				&DA				&DAZ\\
Spectroscopic Data				&WHT			&SDSS			&WHT			&SDSS			&SDSS			&WHT			&SDSS+WHT\\
$\alpha$ (J2000; h\,m\,s)			&09 59 04.69		&11 59 33.10		&12 21 50.81		&12 47 40.93		&13 20 44.68 		&15 06 26.18		&15 57 20.77\\
$\delta$ (J2000; $\degr \ ' \ ''$)		&$-$02 00 47.6		&$+$13 00 31.6	&$+$12 45 13.3	&$+$10 35 56.1	&$+$00 18 54.9 	&$+$06 38 45.9	&$+$09 16 24.6\\
FUV (0.15\,$\mu$m; AB\,mag)		&...				&$>20$			&$19.55\pm0.06$	&$18.23\pm0.02$	&$16.52\pm0.03$	&$18.40\pm0.02$	&$18.03\pm0.02$\\
NUV (0.23\,$\mu$m; AB\,mag)		&...				&$19.62\pm0.09$	&$18.78\pm0.02$	&$18.33\pm0.01$	&$16.76\pm0.01$	&$17.31\pm0.01$	&$18.16\pm0.01$\\
$u$ (0.36\,$\mu$m; AB\,mag)		&$18.52\pm0.02$	&$18.22\pm0.03$	&$18.54\pm0.02$	&$18.63\pm0.02$	&$17.08\pm0.02$	&$16.96\pm0.01$	&$18.50\pm0.02$\\
$g$ (0.47\,$\mu$m; AB\,mag)		&$18.09\pm0.03$	&$18.14\pm0.02$	&$18.19\pm0.02$	&$18.50\pm0.01$	&$17.13\pm0.01$	&$16.57\pm0.01$	&$18.45\pm0.01$\\
$r$ (0.62\,$\mu$m; AB\,mag)		&$18.36\pm0.02$	&$17.75\pm0.01$	&$18.40\pm0.02$	&$18.82\pm0.02$	&$17.47\pm0.02$	&$16.63\pm0.01$	&$18.81\pm0.01$\\
$i$ (0.75\,$\mu$m; AB\,mag)		&$18.54\pm0.02$	&$17.70\pm0.02$	&$18.56\pm0.02$	&$19.13\pm0.02$	&$17.72\pm0.02$	&$16.79\pm0.01$	&$19.13\pm0.01$\\
$z$ (0.89\,$\mu$m; AB\,mag)		&$18.88\pm0.06$	&$17.82\pm0.02$	&$18.86\pm0.05$	&$19.33\pm0.06$	&$18.05\pm0.03$	&$16.93\pm0.01$	&$19.34\pm0.04$\\
$J$ (1.25\,$\mu$m; mag)			&$18.32\pm0.06$	&$17.40\pm0.03$	&$18.43\pm0.07$	&$18.95\pm0.08$	&$17.62\pm0.04$	&$16.41\pm0.01$	&$18.82\pm0.06$\\	
$H$ (1.64\,$\mu$m; mag)			&$18.17\pm0.09$	&$17.32\pm0.06$	&$18.39\pm0.10$	&$18.90\pm0.11$	&$17.71\pm0.08$	&$16.37\pm0.02$	&$19.03\pm0.14$\\
$K$ (2.20\,$\mu$m; mag)			&$17.62\pm0.11$	&$17.08\pm0.08$	&$18.01\pm0.15$	&$18.47\pm0.20$	&$17.53\pm0.11$	&$16.35\pm0.03$	&$18.35\pm0.15$\\
IRAC1 (3.55\,$\mu$m; $\mu$Jy)	&$80\pm4$		&$42\pm2$		&$40\pm5$		&$25\pm2$		&$35\pm2$		&$75\pm4$	 	&$35\pm2$\\
IRAC2 (4.49\,$\mu$m; $\mu$Jy)	&$76\pm4$		&$28\pm2$		&$40\pm6$		&$17\pm2$		&$25\pm2$		&$50\pm3$	 	&$40\pm2$\\

$T_{\rm eff}$ (K)				&$13280\pm20$	&$9500\pm500$	&$12250\pm20$	&$18540\pm150$	&$20220\pm180$	&$10670\pm10$	&$22810\pm40$\\
$\log\,g$ (cm\,s$^{-2}$)			&$8.06\pm0.03$	&8.0				&$8.20\pm0.03$	&$7.86\pm0.03$	&$8.37\pm0.04$	&$8.21\pm0.03$	&$7.54\pm0.01$\\
$M$ ($M_{\odot}$)				&$0.64\pm0.02$	&0.59			&$0.73\pm0.02$	&$0.54\pm0.02$	&$0.86\pm0.02$	&$0.73\pm0.02$	&$0.42\pm0.01$\\
$d$ (pc)						&$203\pm4$		&$110\pm00$		&$180\pm4$		&$396\pm9$		&$150\pm4$		&$71\pm2$		&$566\pm2$\\
IR Excess$^b$					&Dust			&Atm			&Dust			&dM/L			&Bkgd			&None			&Dust\\	

\hline

\end{tabular}
\end{center}

\flushleft
{\em Note}.  Ultraviolet, optical, and near-infrared photometry are from {\em GALEX} \citep{mar05}, 
SDSS \citep{aba09}, and UKIDSS \citep{law07} respectively.   For stars with multiple spectroscopic or 
photometric datasets, table values are the weighted average of available measurements.  The SDSS 
photometry is given in PSF magnitudes; the only catalog entries appropriate for point sources.  Stellar 
parameters are derived by fitting the Balmer line profiles as described in \S3.
\smallskip

$^a$This star was selected independently from the six primary science targets (see \S3.2).
\smallskip

$^b$See \S3 for the various infrared excess descriptions.

\end{table*}

Another way in which near-infrared photometry of white dwarfs is advantageous is that circumstellar dust 
orbiting within the Roche limit of the stellar remnant becomes heated sufficiently to emit in this wavelength 
range \citep{kil06}.  This potential is precluded for main-sequence stars because the analogous spatial region, 
where km-size or larger solid bodies are tidally destroyed, does not extend significantly above their surfaces 
\citep{dav99}.  Furthermore, any material generated in that narrow region above the star will rapidly dissipate 
due to radiation pressure and drag forces.  Thus, while asteroids and comets commonly pass sufficiently close 
to the Sun, and presumably other main-sequence stars that host planetary systems, the tidal disruption and 
subsequent accretion of this debris can only be witnessed at white dwarfs (and possibly neutron stars; 
\citealt{wan06}).

This paper reports warm {\em Spitzer} IRAC measurements and optical spectroscopy of six white dwarf 
candidates selected for apparent $K$-band excess fluxes based on the ground-based photometric surveys 
the Sloan Digital Sky Survey (SDSS) and the UKIRT Infrared Deep Sky Survey (UKIDSS).  Target selection 
criteria and survey data are given in \S2, along with a description of the {\em Spitzer} and ground-based
observations.  The spectroscopic and photometric data analysis is presented in \S3, with a summary of
the overall results in \S4.

\section{TARGET SELECTION AND OBSERVATIONS}

The six primary science targets were selected as described in detail by \citet{gir11}.  Briefly summarizing, DA 
white dwarf candidates were selected both photometrically and spectroscopically from within SDSS DR7 
\citep{aba09}, and spatially cross-correlated with sources in the UKIDSS \citep{law07} DR4 (and later DR8).  
A $ugriz$ color selection was implemented based on the \citet{eis06} SDSS DR4 white dwarf catalog, which 
was then used to select and download spectra for 7444, and photometry (only) for 9341, DA star candidates 
with $g<19$\,AB\,mag.  Cross-correlation with UKIDSS DR8 then resulted in 1884 objects in common with 
both $H$- and $K$-band detections; these were all fitted with white dwarf atmospheric models to probe for 
$K$-band photometric excess.  Of these 1884 cross-correlated sources, 147 objects were identified as 
having an infrared excess, including 12 objects (7 spectroscopic and 5 photometric only) in which the 
excess was most pronounced in the $K$-band, and compatible with either a brown dwarf later than L7
or a dust disk; the UKIDSS data alone could not typically distinguish between those two possibilities.

Six disk candidate white dwarfs found in this manner were selected as targets for {\em Spitzer} Cycle 6 
observations, and are listed in Table \ref{tbl1}.  Two of these targets (1320 and 1557) were independently 
identified by \citet{ste11}.  They performed a cross-correlation of the white dwarf catalogs of \citet{eis06} and 
\citet{mcc99} with UKIDSS DR8 to identify stars with excess emission via optical and infrared colors, as well 
as spectral fitting.  A seventh target, the DQ peculiar white dwarf LP\,494-12 (SDSS\,1159; \citealt{eis06}) 
was selected on the basis of an independent measurement of $K$-band excess \citep{far09a}, corroborated 
by its UKIDSS photometry.

\subsection{Optical Spectroscopy}

\begin{figure}
\includegraphics[height=86mm,angle=-90]{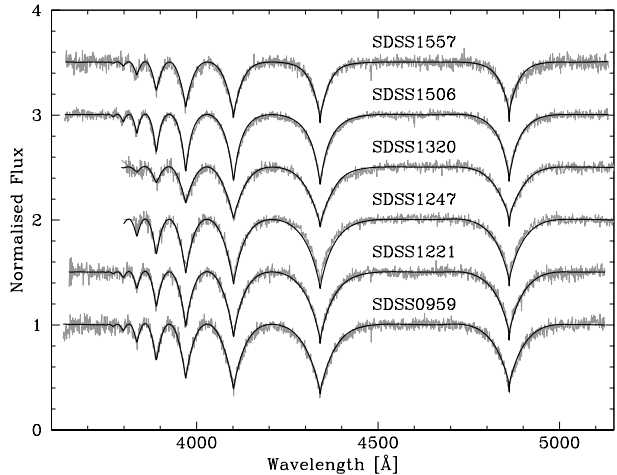}
\caption{Normalized SDSS and WHT spectra of the six DA white dwarfs in the study are shown in gray, along 
with the best fit models plotted in black.  The SDSS spectra stop at 3800\,\AA.  The WHT spectra of 0959, 1221, 
and 1557 each contain a narrow absorption line near 4481\,\AA, that is due to Mg\,{\sc ii}. 
\label{fig1}}
\end{figure}

Of the six primary science targets, SDSS spectra were available within SDSS DR7 for 1247, 1320, and 
1557.  In addition, long slit spectroscopy was obtained for 0959, 1221, 1506, and 1557 during two observing 
runs in 2010 April and May using using ISIS, the dual beam spectrograph mounted on the 4.2\,m William 
Herschel Telescope (WHT).  The primary aim of these observations was to corroborate the photometrically 
selected DA candidates (0959, 1221, 1506) and to improve on the SDSS spectroscopy of 1557.  For this 
purpose, the R600 grating was used in the blue arm with a $1''$ slit and a 2 (spectral) by 3 (spatial) pixel 
binning.  This setup covered the wavelength range $3650 - 5120$\,\AA, i.e., the entire Balmer series except 
H$\alpha$, with an average dispersion of 0.88\,\AA\ per binned pixel in the blue arm, and $7691-9184$\,\AA\ 
with 0.99\,\AA\ per binned pixel in the red arm.  The red spectra had relatively low signal-to-noise ratio (S/N).

The blue spectra were de-biased and flat-fielded using the {\sc starlink}\footnote{Developed and maintained 
by the Joint Astronomy Centre and available from http://starlink.jach.hawaii.edu/starlink} packages {\sc kappa} 
and {\sc figaro} and then optimally extracted using the {\sc pamela}\footnote{Developed and maintained by T. R. 
Marsh and available from http://www.warwick.ac.uk/go/trmarsh} code \citep{mar89}.  The extracted spectra were 
wavelength calibrated using CuNe and CuAr arc lamp exposures, and finally flux calibrated using observations 
of appropriate standard stars obtained with the same instrumental setup.  The photometric classification of 1221, 
1506, and 1557 as hydrogen dominated white dwarfs was confirmed by the ISIS spectroscopy.  Normalized WHT 
and SDSS spectra of all {\em Spitzer} DA targets are shown in Figure \ref{fig1}. 

Additional observations of 0959 and 1320 were obtained with the VLT using X-Shooter \citep{dod06} in service 
mode in 2010 June-July and 2011 February-March.  The raw frames were reduced using the X-Shooter pipeline 
version 1.3.7 within {\sc gasgano}\footnote{http://www.eso.org/sci/software/gasgano}.  The standard recipes were 
used with default settings to extract and wavelength calibrate each spectrum.  The final extraction of the science 
and spectrophotometric standard spectra was carried out using {\sc apall} within IRAF.  Finally, the instrumental 
response was removed by dividing each observation by the response function, calculated by dividing the 
associated standard by its corresponding flux table. 

\begin{figure*}
\includegraphics[width=172mm]{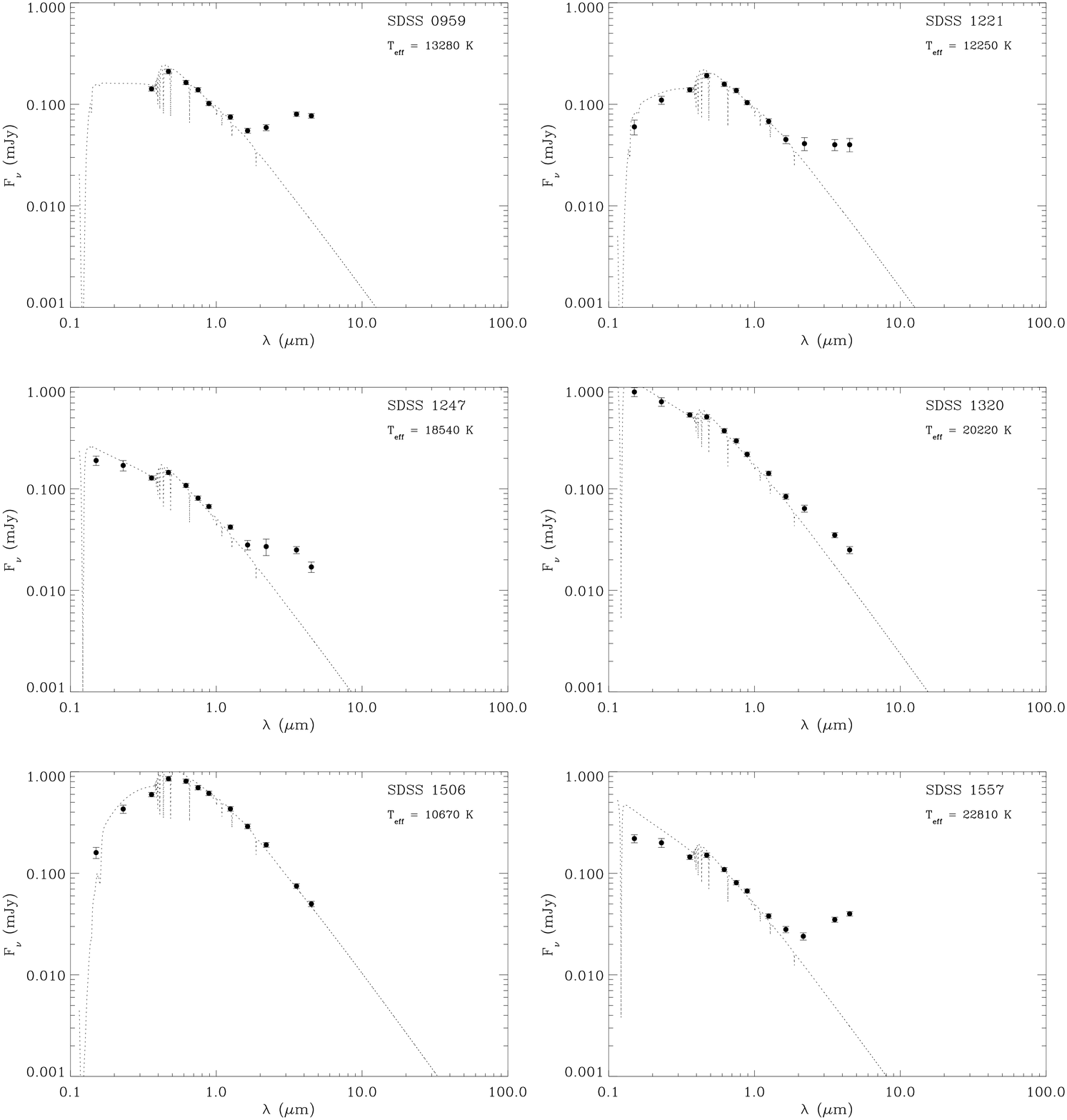}
\caption{Ultraviolet through infrared spectral energy distributions of the DA-type, infrared excess candidates.  
Stellar atmosphere models are plotted as dotted lines, using parameters derived from model fits to hydrogen 
Balmer lines in the SDSS or WHT spectra, and matched to the observed $g$-band fluxes.  Table \ref{tbl1} 
photometry is shown as data points with error bars.
\label{fig2}}
\end{figure*}

\subsection{Additional Near-Infrared Photometry}

\begin{table}
\begin{center}
\caption{Independent $JHK$ Photometry\label{tbl2}} 
\begin{tabular}{@{}lccc@{}}
\hline\hline

Star			&$J$				&$H$			&$K_s$\\
			&(mag)			&(mag)			&(mag)\\

\hline

1221$^a$		&$18.49\pm0.05$	&$18.21\pm0.05$	&$18.05\pm0.05$\\
1221$^b$		&$18.50\pm0.05$	&$18.27\pm0.05$	&$18.22\pm0.10$\\
\\
1320$^a$		&$17.64\pm0.05$	&$17.44\pm0.05$	&$17.49\pm0.05$\\
1320$^b$		&$17.60\pm0.05$	&$17.61\pm0.07$	&$17.58\pm0.11$\\
\\
1557			&$19.05\pm0.05$	&$18.92\pm0.05$	&$18.56\pm0.05$\\

\hline

\end{tabular}
\end{center}

$^a$Aperture photometry
\smallskip

$^b$PSF-fitting photometry

\end{table}

Independent near-infrared photometry for 1221, 1320, and 1557 was obtained on the 23 March 2011 
with the WHT using the Long-Slit Intermediate Resolution Infrared Spectrograph (LIRIS; \citealt{man98}).  
Images taken in a 9-point dither pattern were obtained in the $J$-, $H$-, and $K_s$-band filters (MKO
system; R. Karjalainen 2011, private communication) with typical total exposure times of 270\,s in clear 
conditions.  Three standard star (ARNICA; \citealt{hun98}) fields were observed in a similar manner for 
photometric zero-point calibration.  The data were reduced in the standard manner, by subtracting a 
median sky from each image in the dithered stack, flat-fielding (using sky flats), then averaging and 
recombining frames.

LIRIS suffers from what is known as a detector reset anomaly, which appears in certain frames as a 
discontinuous jump (in dark current) between the upper and the lower two quadrants. To remove this 
unwanted signal, after flat-fielding and sky subtraction, the detector rows were collapsed into a median 
column (with real sources rejected), and subsequently subtracted from the entire two dimensional image.  
The resulting fully reduced frames exhibit smooth backgrounds, free of the anomalous gradient. 

Aperture photometry of standard stars was performed using $r=3\farcs75$ aperture radii and sky annuli 
of $5''-7\farcs5$ in size.  For the relatively faint science targets, smaller photometric apertures were employed
with corrections derived from several brighter stars within the same image field and filter.  For both 1221 and 
1320, point-spread function (PSF) fitting (i.e., {\sc daophot}) was used in addition to photometry with small 
apertures.  All data taken in the $K_s$-band filter were flux-calibrated using the ARNICA $K$-band standard 
star photometry, and the error introduced by this should be significantly smaller than the 5\% absolute 
calibration uncertainty.  The independent $JHK_s$ photometry is listed in Table \ref{tbl2}.

\subsection{{\em Spitzer} IRAC Observations}

Near-infrared imaging observations of the white dwarf targets were obtained with the warm {\em Spitzer 
Space Telescope} \citep{wer04} during Cycle 6 using the Infrared Array Camera (IRAC; \citealt{faz04}) at 
3.6 and 4.5\,$\mu$m.  The total integration time in each channel was 1200\,s, where the observations 
consisted of 40 frames taken in the cycling (medium) dither pattern with 30\,s individual exposures.  All 
images were analyzed and fluxes measured as in \citet{far10b} using $0.6$\,pixel$^{-1}$ mosaics created
using MOPEX.  In cases where the flux of a neighboring source was a potential contaminant of the white 
dwarf photometry (i.e.,\ 1221 and 1247, see \S3.1), steps were taken to minimize or remove any such 
external contributions, including small aperture photometry and PSF fitting with {\sc daophot} and {\sc 
apex}.

Of all IRAC targets, only 1506 is detected in the {\em WISE} Preliminary Data Release.  Reliable data 
exists only at 3.4\,$\mu$m, where the flux ($76\pm8\,\mu$Jy) is in agreement with the IRAC photometry, 
albeit with larger errors.  Generally, the science targets are too faint for {\em WISE}, where confusion can 
be a serious concern (see \citealt{mel11}), and hence the choice of warm {\em Spitzer}.

\section{RESULTS AND ANALYSIS}

The SDSS and WHT spectra of the six DA white dwarfs were fitted using model atmosphere spectra 
computed with the code described by \citet{koe10}, and following the method described by \citet{reb07}. 
In brief, the grid of model spectra was fitted to the normalized Balmer line profiles, leading to a `hot' and 
`cold' solution of roughly equal absorption line equivalent width.  This degeneracy is broken by fitting the 
models also to the slope of the flux-calibrated spectra.  The best-fit models are overplotted on the data in 
Figures \ref{fig1} and \ref{fig2}, and the parameters are reported in Table \ref{tbl1}.

All targets stars but one reveal excess emission at IRAC wavelengths when compared to the atmospheric 
models, derived from their optical spectroscopy.  However, there are only four genuine cases where the 
excess radiation is associated with an additional component physically associated with the system: three 
dust disks and one (likely) low-mass stellar companion.  All individual stars are discussed below in some 
detail.

\begin{figure}
\includegraphics[height=86mm,angle=-90]{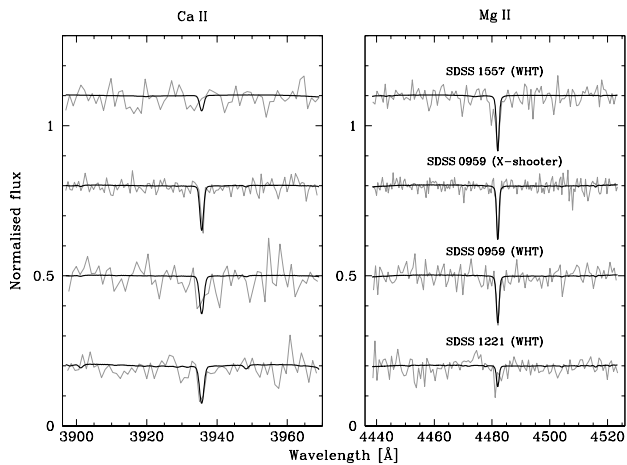}
\caption{Black lines are model fits to the Ca\,{\sc ii} 3934\,\AA \ and Mg\,{\sc ii} 4481\,\AA \ line regions in the 
normalized ISIS and X-Shooter spectra, plotted in gray, of all three white dwarfs found to have circumstellar 
dust.  The stars are displayed top-to-bottom by decreasing effective temperature: 22\,810\,K, 13\,280\,K, 
12\,250\,K.
\label{fig3}}
\end{figure}

\subsection{New Disk-Polluted White Dwarfs}

Three stars -- 0959, 1221, 1557 -- display infrared fluxes consistent with dust disks orbiting within their 
respective Roche limits for large asteroids.  The excess emissions are too strong, and the infrared colors 
are incorrect for substellar or planetary companions \citep{far09b,far08a}.  Specifically, even for the case 
of a substellar object with anomalously red colors, the 4.5\,$\mu$ flux requires its size to be several to ten 
times that of Jupiter, in contrast with observations and models of brown dwarfs and planets \cite{leg10}.
Firmly corroborating the disk interpretation, each of these DA white dwarfs was found to be metal-lined 
in follow up WHT and VLT optical spectroscopy, consistent with ongoing infall of circumstellar material.  
Specifically, their ISIS spectra contain narrow absorption lines near 3934\,\AA\ and 4481\,\AA, which are 
due to Ca\,{\sc ii}\,K and Mg\,{\sc ii}, respectively; they are DAZ stars.  Figure \ref{fig3} displays model fits 
to the the detected metal absorption features, using {\sc tlusty} and {\sc synspec} \citep{hub95,lan95} in 
order to estimate the photospheric abundances.

Only Mg and Ca are detected, and Table \ref{tbl3} reports the determined abundances and upper limits for 
these two elements.  The Mg/Ca ratio observed in 0959 is significantly larger than that found in chondrites, 
but similar to that found for the highly polluted white dwarf GALEX\,1931 \citep{ven11}, where the relative 
poverty of Ca, Ti, and Al is potentially consistent with the removal of crust and mantle in a differentiated,
terrestrial-like body during the post-main sequence evolution of the star \citep{mel11}.  The Mg abundances 
determined from the X-Shooter and ISIS spectra of 0959 are consistent to within the errors, while the low S/N 
around the Ca\,{\sc ii}\,K line prevents a meaningful abundance estimate for that element based on the ISIS
data.  For 1221 and 1557, only upper limits are derived for Mg and Ca respectively, partly due to the inferior 
quality of the ISIS spectra compared to the X-Shooter data.  However, another factor is that the undetected 
atomic transitions are not easily excited at the respective effective temperature of these stars \citep{zuc03}.

For all three stars, the excess infrared emissions are fitted with optically thick, flat disk models \citep{jur03}.
As discussed in previous white dwarf-disk studies, optically thin dust models are inferior, primarily due to 
the Poynting-Robertson timescales involved \citep{von07}.  Not only would such dust be removed on the 
timescale of years, but the particles orbit once every few hours, ensuring the disk will relax rapidly into a 
flat configuration; a geometrically thin, optically thick disk circumvents these issues.  Importantly, such a disk 
can harbor sufficient dust mass to account for the most highly polluted helium atmosphere stars \citep{far10a}, 
whereas an optically thin disk cannot \citep{jur09}.

\begin{table}
\begin{center}
\caption{Relative Number Abundances in Disk-Polluted Stars\label{tbl3}} 
\begin{tabular}{@{}lrrr@{}}
\hline\hline

Object		&$\log$\,(Mg/H)	&$\log$\,(Ca/H)		&Mg/Ca\\

\hline

Sun			&$-4.45$			&$-5.66$			&16\\
Chondrites	&$-0.72$			&$-1.96$			&17\\
0959			&$-5.2$			&$-7.0$			&59\\
1221			&$<-5.6$			&$-7.5$			&$<85$\\
1557			&$-4.5$			&$<-5.7$			&$>15$\\

\hline

\end{tabular}
\end{center}

{\em Note}.  Measured abundance uncertainties are 0.2\,dex.

\end{table}

\begin{figure}
\includegraphics[width=86mm]{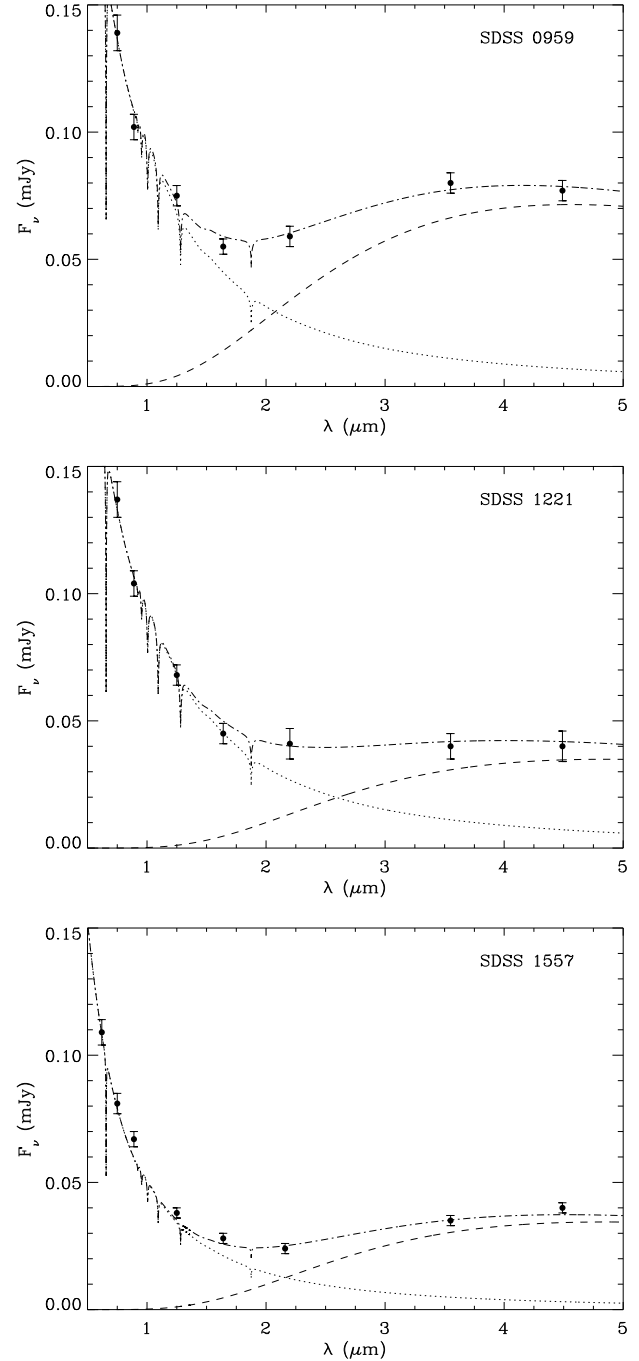}
\caption{Infrared excesses fit by circumstellar dust disk models whose parameters are listed in Table 
\ref{tbl4}.  The plot features are the same as in Figure \ref{fig2}, but on a linear scale to emphasize the 
infrared excesses.  Dashed lines represent emission from optically thick, flat disk models, and a the
dashed-dotted lines represent the sum of the stellar and circumstellar model fluxes. 
\label{fig4}}
\end{figure}

The fitted disk parameters are listed in Table \ref{tbl4}, and the model fluxes are shown together with the 
observational data in Figure \ref{fig4}.  For each star, $R/d$ is derived from the atmospheric models using 
the spectroscopically derived stellar parameters and observed photometry: $R$ is determined from $T_{\rm 
eff}$ and $\log\,g$ while $d$ is specified by $m-M$ for the same (e.g., $g-M_g$, $r-M_r$).  For fixed $R/d$, 
there are three free parameters for a flat disk; inner disk edge temperature, outer disk edge temperature, 
and inclination.  These models contain a modest amount of degeneracy in their ability to fit the data, even 
when longer wavelength data, such as 8 and 24\,$\mu$m photometry are available \citep{jur07}. However, 
the models provide good, if broad constraints on the geometry and temperature of the disks.  The inner disk
edge is fairly well constrained by the excess emission at 2.2 and 3.6\,$\mu$m, but not uniquely so, while the 
4.5\,$\mu$m flux can be reproduced by a relatively wide temperature range and a higher inclination, or a 
narrower temperature range and a lower inclination.  This is because the total disk emission is a function 
of its temperature profile and solid angle, and the latter is comprised of the inclination and the ring annuli. 

Despite this uncertainty, the infrared excesses cannot be satisfactorily fit with inner disk temperatures
below 1200\,K, nor with significant amounts of $T<500$\,K dust.  These new discoveries essentially mimic
the thermal emission profiles observed at 16 dusty white dwarfs with longer wavelength data in addition to 
the wavelengths covered by warm {\em Spitzer} \citep{far10b}.  The excess at 0959 is sufficiently strong that it 
requires a nearly face-on disk, while the disk inclinations for 1221 and 1557 are less constrained.  Importantly, 
the excesses are consistent with inner disk temperatures approaching or exceeding that which causes rapid
sublimation of typical dust grains; gas thus produced gives rise to viscous drag and enhances inflow of disk 
material onto the stellar surface \citep{raf11}.  While the outer disk temperatures are not more tightly bound, 
the range of acceptable values places all the debris within $1.5\,R_{\odot}$, where large, solid bodies will
be shredded.  Overall, the data are consistent with circumstellar debris originating in large, star-grazing 
solid bodies that is now being gradually falling onto the stellar surface.

\begin{table}
\begin{center}
\caption{Disk Parameters\label{tbl4}} 
\begin{tabular}{@{}cccccccc@{}}
\hline\hline

Star			&$T_{\rm eff}$	&$R/d$		&$T_{\rm in}$	&$T_{\rm out}^a$	&$r_{\rm in}$	&$r_{\rm out}^a$	&$\cos{i}$\\
			&(K)			&($10^{-12}$)	&(K)			&(K)				&($R_*$)		&($R_*$)			&\\

\hline

0959			&13280		&1.37		&1600		&800			&10			&25				&1.0\\
1221			&12250		&1.41		&1400		&800			&11			&23				&0.7\\
1557			&22810		&0.73		&1400		&800			&25			&52				&0.5\\

\hline

\end{tabular}
\end{center}

$^a$Owing to the wavelength coverage, the outer disk temperature and radius is not as well-constrained 
by the data as the inner disk temperature.  See \S3.1 for a detailed discussion of uncertainties in the disk 
parameters.

\end{table}

\subsubsection{0959}

\begin{figure}
\includegraphics[height=86mm,angle=-90]{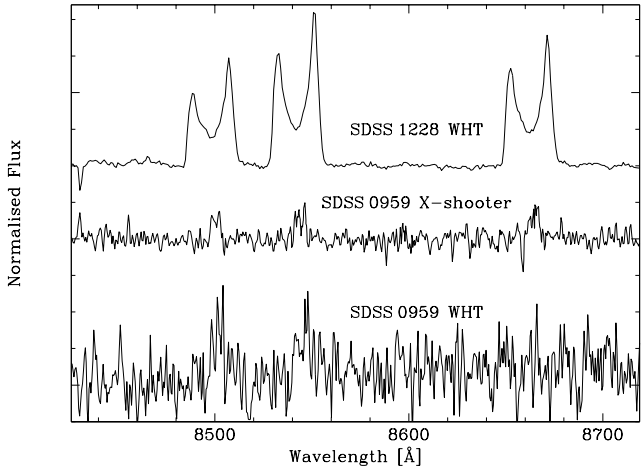}
\caption{The Ca\,{\sc ii} triplet region of 0959 in the normalized WHT and X-Shooter spectra exhibit 
clear emission lines from a gaseous disk component.  For comparison, also plotted is the normalized 
spectrum of the prototype gas disk around SDSS\,1228 \citep{gan06}.  The narrower emission features 
at 0959 indicate a disk that is significantly inclined relative to that seen at 1228, and consistent with the 
high fractional infrared luminosity measured with IRAC. 
\label{fig5}}
\end{figure}

This white dwarf displays Ca\,{\sc ii} emission lines that are the hallmark of closely-orbiting metallic 
gas.  These emission features are clearly detected in the X-shooter data, and, upon close inspection, 
are marginally visible in the noisy red ISIS spectrum.  Both red spectra are shown in Figure \ref{fig5}, 
along with SDSS\,1228, the prototypical white dwarf with emitting metal gas and dust \citep{bri09,gan06}.  
Notably, the Ca\,{\sc ii} line profiles for this star are substantially narrower and weaker compared to those 
detected in the other three published cases.  In 0959 the full width at zero intensity and equivalent width are 
close to 250\,km\,s$^{-1}$ and 3\,\AA, respectively, while the known cases have values in the range $1000-
1400$\,km\,s$^{-1}$ and $10-60$\,\AA \ \citep{mel10,gan08,gan07,gan06}.  Assuming that the structure of the
gas disks in 0959 and 1228 are similar, the velocity width of the Ca\,{\sc ii} lines in the two systems can be
used to estimate the inclination of the disk orbiting 0959.  For 1228, two independent estimates yield an 
inclination of $70\degr$ \citep{mel10,bri09}, and this yields $i\approx12\degr$ for 0959. The independently 
estimated disk inclination from the model fitted to the infrared emission is 10\degr, consistent with a near 
face-on disk of dust and gas.  \citet{gan06} found that the Ca\,{\sc ii} emission lines in 1228 are optically thick, 
and hence a low disk inclination will naturally result also in a lower equivalent width.  

SDSS\,0959 becomes the fourth published case of a white dwarf with detectable, gaseous debris in addition 
to solid dust particles in a circumstellar disk \citep{mel10,far10b, bri09}.  At $T_{\rm eff}=13\,300\,$K it is by far 
the coolest star to host such dual-phase debris, demonstrating conclusively that the {\em detected} gas in dust 
disks is {\em not} related to grain sublimation in the presence of higher temperature white dwarfs.  In the three 
well-studied cases, the gaseous and solid debris are essentially spatially coincident \citep{mel10,bri09}, both 
spanning a region from around 20 to 100 stellar radii \citep{gan06}.  Critically, there is a distinct lack of emission 
from gas in the innermost regions where sublimation is expected, and which may be due to lower disk surface 
density, resulting from enhanced viscosity and inward drift of gas \citep{mel10}.  Corroborating this picture, 
sublimation of dust at 0959 is expected only within 15 stellar radii, and hence cannot account for emitting gas
out to 100 radii.

It is uncertain if the current disk at 0959 and other gas-dust disk hosts are recently created or extant disks that
have been impacted by a additional body or material.  Regardless, the presence of detectable gas is almost 
certainly related to the destruction of solid material.

\subsubsection{1221}

The IRAC images of this star reveal a point-like source neighboring the white dwarf.  Both sources 
were simultaneously modeled and deconvolved with PSF-fitting photometry.  The neighbor is faintly but 
clearly visible in the UKIDSS $H$-band image, and is identified as a separate source in that survey.  The 
LIRIS $K_s$-band images place it $1\farcs8$ distant distant at position angle $256\degr$.  Together with its 
IRAC fluxes, the $JHK$ photometry of the neighbor yields colors inconsistent with a (substellar) companion 
\citep{pat06}.  Although relatively faint, it appears to be diffuse and extended in the ground-based images 
and is thus a background galaxy.  Both the UKIDSS and LIRIS photometry of the white dwarf reveal strong 
infrared excess at $K$, with the latter dataset confidently free of photometric contamination by the neighbor, 
in agreement with the excess determined from IRAC photometry.  No Ca\,{\sc ii} lines are detected in the 
low S/N red ISIS spectrum.

\subsubsection{1557}

The mass of the white dwarf in 1557 is significantly below the average mass of field white dwarfs \citep{lie05}. 
While the fit to the SDSS spectrum is consistent with a low-mass, carbon-oxygen core, the WHT data are more 
suggestive of a helium core. The SDSS data are of relatively poor quality at S/N $\approx10$, while the WHT 
spectrum has S/N $\approx30$.  Table \ref{tbl1} lists the weighted average of these two solutions, favoring the 
low mass interpretation.  If a low surface gravity is corroborated in future observations, this may signify an 
innermost planet that was consumed during the first ascent giant phase, ejecting sufficient envelope mass to 
prevent the onset of helium ignition \citep{nel98}.  This picture is consistent with the evidence presented here 
for a remnant planetary system.  As in the case of 1221, no Ca\,{\sc ii} emission is seen in the poor quality ISIS
data.

\subsection{Unusual Near-Infrared Colors for 1159}

\begin{figure}
\includegraphics[width=86mm]{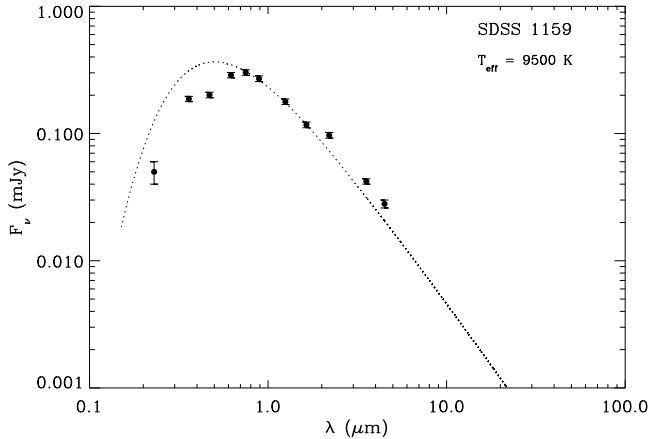}
\caption{Spectral energy distribution of the DQ peculiar star LP\,494-12, which has $H-K>0.25$\,mag in both 
UKIDSS and independently obtained photometry.  Fitted to the $izJH$ photometry is a 9500\,K stellar model 
for a pure helium atmosphere white dwarf (see \S3.2).
\label{fig6}}
\end{figure}

This DQ peculiar star (LP\,494-12, WD\,1156$+$132) was selected on the basis of independent $JHK$ 
photometry that revealed a 0.3\,mag excess at $K$ band \citep{far09a} relative to model expectations for 
a 10\,000\,K, helium atmosphere white dwarf \citep{hol06,fon01}; its UKIDSS photometry independently 
corroborates this result.  Due to the ${\rm C}_2$ absorption features at blue-green wavelengths, a pure 
helium atmospheric model was fitted only to the $izJH$ photometric data, as shown in Figure \ref{fig6}.  
The IRAC fluxes reveal only a Rayleigh-Jeans tail to the spectral energy distribution, but at a shifted, 
lower temperature relative to the $0.6-1.5$\,$\mu$m continuum.  As found for another DQp star (LHS\,2293; 
\citealt{far09a}), this apparent excess is likely a re-distribution of emergent flux due to the highly absorbed 
regions within the atmospheric ${\rm C}_2$ bands.  Thus the infrared excess only exists relative to a pure 
helium atmosphere, and is likely due to a distinct, absorption-induced shape in the energy distribution of 
the star.

\subsection{A Likely Ultracool Companion at 1247}

The infrared excess observed at this star is distinct, as the 3.6\,$\mu$m flux excess is strong, but the 
4.5\,$\mu$m flux is consistent with a Rayleigh-Jeans type slope (see Figure \ref{fig2}).  At first glance this 
may seem puzzling, but most M dwarfs approach this spectral behavior at IRAC wavelengths \citep{pat06,
cus05}.  Neither the UKIDSS nor the IRAC images indicate an additional source (i.e., a spatially detectable 
companion), implying any secondary would have a projected separation smaller than around 70\,AU, and
essentially ruling out a background object as the cause of the observed emission.  While the infrared excess 
is solid, and possibly apparent at $J$ band, the S/N of the UKIDSS photometry is below 10 at $H$ and $K$, 
making it hard to estimate a companion spectral type based on these data.  Assuming the observed excess
emanates from a companion at the correct distance, its 2.2, 3.6, and 4.5\,$\mu$m absolute brightnesses are
11.2, 10.0, and 9.8\,mag.  Comparing these values to the empirical absolute magnitudes of ultracool dwarfs 
at these wavelengths yields an approximate spectral type of L$0\pm1$ \citep{leg10,pat06,vrb04,dah02,kir94}
for a low mass companion.  Superior $JHK$ photometry would tighten up this estimate, and possibly indicate 
a new L-type companion to a white dwarf, which are rare \citep{ste11,far05}.  Radial velocity monitoring of the
white dwarf has the potential to detect variability if the system is a close binary, whereas high angular resolution
imaging may be required if the system has a moderately wide separation.

\subsection{A Likely Background Galaxy at 1320}

The IRAC images of this star have full widths at half maximum that are inconsistent with a single point source; 
they are enlarged by about 20\% in both channels and there is modest elongation in the 3.6\,$\mu$m image 
roughly along P.A. $45\degr$.  Nothing is seen in the UKIDSS images, but a second source appears faintly in 
the LIRIS $H$ and $K_s$-band images at a position consistent with the IRAC data.  The image of this additional 
source is diffuse and extended and almost certainly a background galaxy.  Moreover, there is an offset between 
the image centroids of the science target in the LIRIS and IRAC images, indicating that the IRAC source (and 
hence flux) is coincident with the additional source imaged with LIRIS, and not the white dwarf.  The flux from
this likely galaxy falls within $1\farcs0$ of the white dwarf, and contaminates all available infrared photometry.  
Even with PSF-fitting photometry on the LIRIS images, the resulting colors are slightly too red for a 20\,200\,K 
white dwarf.  Given the overlapping proximity of this source to the white dwarf, the infrared excess is probably 
due entirely to this background object.

\subsection{No Near-Infrared Excess for 1506}

Somewhat ironically, the only star without a corroborated infrared excess via IRAC photometry is the brightest 
source in the sample, and thus the star with the {\em highest} S/N UKIDSS photometry.  However, this star was
selected for the {\em Spitzer} program when the evaluation criteria were still tied to UKIDSS DR4, and has since 
dropped out of the selection based on the release of DR8 and an improved analysis \citep{gir11}.  The model
for 1506 shown in Figure \ref{fig2} shows that an excess is neither observed with {\em Spitzer}, nor in $K$-band.
While this white dwarf was observed with ISIS, confirming its nature as a DA star, it is hence dropped from further
analysis.

\section{DISCUSSION AND CONCLUSIONS}

\subsection{An Unbiased Dust Disk Frequency}

What makes this study so far unique is that previous IRAC programs have either targeted known metal-rich 
white dwarfs, or near-infrared bright white dwarfs.  While the first approach has been highly successful and 
accounts for the bulk of white dwarf disk discoveries to date \citep{far10b,far09b,jur07}, two drawbacks exist.  
First, it requires 8\,m class telescope time using high-resolution spectroscopy to efficiently identify (weak) 
metal lines in white dwarfs \citep{koe05,zuc03}, and second, only about 20\% of metal-contaminated white 
dwarfs have disks \citep{far09b}.  The second method, employed by the \citet{mul07} survey of more than 
130 white dwarfs with $K_s<15$\,mag, was unbiased by the presence or absence of atmospheric metals, 
but only two stars with dust were found \citep{von07}, and is hence only 1.5\% efficient by number of targets.  
Here, five candidate DA white dwarfs with $K$-band excess have been selected with a success rate of 60\% 
for circumstellar dust and 80\% for confirmed infrared excesses physically associated with the system.  One 
of five targets has an infrared excess probably due to a background galaxy.

Based on the candidate selection criteria and numbers given in Table 6 of \citet{gir11}, of 1884 candidate 
and confirmed DA white dwarfs with the necessary UKIDSS detections, there are 12 stars whose data are 
consistent with disk-like, $K$-band excesses.  Several years of Spitzer studies have shown that only 50\% 
of dusty white dwarfs have such $K$-band excesses \citep{far10b}, implying that a decent estimate of the disk 
fraction among DA white dwarfs is at least 1.2\%. However, as seen in this study, in some cases the $K$-band 
data can be the result of neighboring background objects, as well as real companions such as low mass stars 
or brown dwarfs.  Thus a more realistic estimate is $3/5$ of this number, implying at least 0.8\% of DA (and 
presumably all) white dwarfs have dust disks (and atmospheric metals), consistent with previous estimates 
\citep{far09b}.  These results only apply to stars with cooling ages less than 1\,Gyr, as this corresponds to the 
approximate cutoff in effective temperature (8000\,K; \citealt{fon01}) for the DA color selection.  However, {\em 
Spitzer} studies have shown a distinct lack of dust disks at older and cooler metal-contaminated stars; only 
G166-58 has an (anomalous) infrared excess and a cooling age beyond 1\,Gyr \citep{far08b}.

\subsection{A Lack of $K$-Band Emitting Dust Disks at Warm to Hot White Dwarfs}

Interestingly, the DA selection criteria is quite sensitive to warm to hot, hydrogen-rich white dwarfs, as the 
Balmer decrement gives these stars unique colors in that wavelength region \citep{gir11}. However, there 
are no candidates for dust among thousands of stars(!).  Either dust	disks	 at white dwarfs with	$T_{\rm eff}
>25\,000$\,K exist yet rarely exhibit $K$-band excess or dust at such stars is itself rare.  The latter possibility
is somewhat contradictory based on the paradigm of planetary systems that are dynamically rejuvenated at 
the onset of the post-main sequence \citep{bon11,bon10,deb02}.  If this apparent dearth of warmer systems
at $K$-band is real, it would likely require a physical mechanism to preclude such emission (such as rapid
removal of warm grains, or long timescales for dust disk spreading toward the star).

\citet{far11} plot the fractional infrared luminosity of all known white dwarfs with dust circa mid-2010 as a 
function of cooling age, revealing a potential trend of increasing dust emission towards cooler stars.  This 
could be a natural consequence of the potential for cooler white dwarfs to host wider disks: as a star cools 
and its luminosity decreases, dust grains may persist closer to the stellar surface prior to sublimation.  Thus, 
any disk extending inward from the Roche limit has the potential to be wider at cooler white dwarfs.  Conversely, 
at increasingly warmer stars, the inner dust disk edge will be physically further from the star as the radius at which 
grains are rapidly sublimated increases (see Table \ref{tbl4}), eventually exceeding the Roche limit for large solid 
bodies \citep{von07}.  While it is uncertain at which precise effective temperature an evolving dust disk might be 
completely sublimated, the results of this work clearly show that disk emission analogous to G29-38 is rare among 
warm and hot white dwarfs.  More sensitive surveys with {\em Spitzer} and {\em WISE} are needed to address this 
matter more comprehensively.

\section*{ACKNOWLEDGMENTS}

The authors thank the referee Sandy Leggett for feedback which improved the quality and clarity of the 
manuscript.  This work is based in part on observations made with the {\em Spitzer Space Telescope}, 
which is operated by the Jet Propulsion Laboratory, California Institute of Technology under a contract 
with NASA.  Some data presented herein are part of the Sloan Digital Sky Survey, which is managed by 
the Astrophysical Research Consortium for the Participating Institutions (http://www.sdss.org/), and the 
UKIRT Infrared Deep Sky Survey.

\label{lastpage}

\end{document}